\documentclass[prd,draft,amsfonts,amssymb,amsmath,showpacs]{revtex4}

\begin{document}

\title{Dynamical Chiral Symmetry Breaking and Superconductivity 
in the Supersymmetric Nambu$-$Jona-Lasinio Model at finite Temperature and Density}
\author{Tadafumi Ohsaku}
\affiliation{Research Center for Nuclear Physics, Osaka University, Ibaraki, Osaka, Japan.}

\date{\today}

%\maketitle

\newcommand{\bmx}{\mbox{\boldmath $x$}}
\newcommand{\bmy}{\mbox{\boldmath $y$}}
\newcommand{\bmk}{\mbox{\boldmath $k$}}
\newcommand{\bmp}{\mbox{\boldmath $p$}}
\newcommand{\bmq}{\mbox{\boldmath $q$}}
\newcommand{\bmP}{\mbox{\boldmath $P$}}  
\newcommand{\kfey}{\ooalign{\hfil/\hfil\crcr$k$}}
\newcommand{\pfey}{\ooalign{\hfil/\hfil\crcr$p$}}
\newcommand{\qfey}{\ooalign{\hfil/\hfil\crcr$q$}}
\newcommand{\Deltafey}{\ooalign{\hfil/\hfil\crcr$\Delta$}}
\newcommand{\nablafey}{\ooalign{\hfil/\hfil\crcr$\nabla$}}
\newcommand{\Dfey}{\ooalign{\hfil/\hfil\crcr$D$}}
\newcommand{\partfey}{\ooalign{\hfil/\hfil\crcr$\partial$}}
\def\sech{\mathop{\rm sech}\nolimits}

\begin{abstract}

We investigate the dynamical chiral symmetry breaking ( DCSB ) and superconductivity
in a supersymmetric model at finite temperature and density.
We employ the ${\cal N}=1$ four-dimensional generalized supersymmetric Nambu$-$Jona-Lasinio model
( ${\cal N}=1$ generalized ${\rm SNJL}_{4}$ ) 
with a chemical potential as the model Lagrangian, and select the gauge freedom as $U(1)$.
In order to realize the DCSB and BCS-type superconductivity in this model, 
we introduce a SUSY soft mass term. 
Under the finite-temperature Matsubara formalism,
the effective potential and the gap equations are derived in the flamework of the large-$N$ expansion.
The finite-density effect in the DCSB is shown by the critical coupling.
The roles of both the boson and fermion sectors in the superconductivity 
are examined by the quasiparticle excitation spectra and the gap equations.

\end{abstract}

\pacs{11.30.Pb, 11.30.Rd, 74.20.Fg}

\maketitle

\section{Introduction}

Until now, the BCS ( Bardeen-Cooper-Schrieffer ) theory of superconductivity 
has influenced a large part of theoretical physics~[1,2,3,4,5,6,7,8].
In the investigations of color superconductivity ( CSC ), 
the BCS method in relativistic field theory is widely used~[6].
Because of its fundamental importance, it is interesting to construct
a supersymmetric theory of superconductivity.

Recently, several attempts to examine the theories of 
CSC from the viewpoint of supersymmetry ( SUSY ) appeared in literature~[9,10].
According as the results of the nonperturbative method of SUSY gauge theories~[11], 
Ref.~[9] discussed the symmetry breaking patterns, especially the breaking of baryon density symmetry 
$U(1)_{B}$ of the ${\cal N}=1$ SQCD ( supersymmetric QCD ) with a nonzero chemical potential $\mu$ and 
a SUSY breaking mass $\Delta$ for squarks.
However, any gauge-symmetry-breaking two-body pairs like diquarks were not treated in Ref.~[9]. 
Furthermore, because the validity of the exact results of SUSY gauge theories to the problem of CSC is not clear, 
the method of Ref.~[9] could apply to the situation of $\mu < \Delta \ll \Lambda_{SQCD}$, 
where $\mu$ and $\Delta$ ( both break SUSY explicitly )
could be regarded as a small perturbation to a SUSY gauge theory.
Under a similar context, the authors of Ref.~[10] proposed a toy model for giving diquark-like condensates.
They introduced an $SO(N)$ gauge interaction between quark superfields stronger than that of $SU(3_{c})$,
and then they argued the $SO(N)$ gauge dynamics gives diquark-type condensations in their model.
This $SO(N)$ gauge interaction seems artificial.

The purpose of this work is different from that of Refs.~[9,10]. 
We investigate the dynamical chiral symmetry breaking ( DCSB ) and superconductivity
in a SUSY condensed matter at $\mu\sim{\cal O}(\Delta)$.
$\mu$ is the characteristic energy scale of a condensed matter system,
while $\Delta$ gives the SUSY breaking scale.
Roles of superpartners in the DCSB and superconductivity might appear clearly at $\mu\sim{\cal O}(\Delta)$.
For the purpose, we intend to make our method parallel with the ordinary BCS theory, 
and supersymmetrize it by using a generalized version of the SUSY Nambu$-$Jona-Lasinio ( SNJL ) 
model~[12,13,14,15,16].
We choose the simplest case of the gauge symmetry, namely $U(1)$, to avoid possible complicated situations
of breaking patterns of gauge and flavor symmetries.
The procedure of this paper will make a first step toward the investigation of SUSY superconductivity
with more large gauge symmetries.
We derive and solve a BCS-type self-consistent gap equation of the SUSY superconductivity,
and examine the thermodynamics.
We discuss characteristic aspects of our SUSY BCS theory by comparing with 
the results of the non-relativistic and relativistic BCS theories~[1,6,17,18].  
To obtain the effective potential at finite temperature, we use the imaginary time Matsubara formalism.
The ${\cal N}=1$ SUSY is explicitly broken at finite temperature.
The SNJL model was first introduced to investigate DCSB in a SUSY field theory~[12,13], 
and it was also applied to the theory of the top condensation of the minimal SUSY standard model~[8]. 
The examination on the phenomenon of the chiral symmetry breaking of SNJL at finite-temperature
and zero-density was given in literature~[15].

\section{Formalism}

The starting point for our investigation is the following Lagrangian of an ${\cal N}=1$ SUSY model
in four-dimensional spacetime:
\begin{eqnarray}
{\cal L} &=& \int d^{2}\theta d^{2}\bar{\theta} \Bigl( \Phi^{\dagger}_{+}e^{V}\Phi_{+}+\Phi^{\dagger}_{-}e^{-V}\Phi_{-}
+\frac{G_{1}}{N}\Phi^{\dagger}_{+}\Phi^{\dagger}_{-}\Phi_{+}\Phi_{-} 
+\frac{G_{2}}{N}\Phi^{\dagger}_{+}\Phi^{\dagger}_{+}\Phi_{+}\Phi_{+} 
+\frac{G_{3}}{N}\Phi^{\dagger}_{-}\Phi^{\dagger}_{-}\Phi_{-}\Phi_{-} \Bigr). 
\end{eqnarray}
Here, $\Phi_{+}$ and $\Phi_{-}$ are chiral matter superfields, $N$ is the number of flavor.
$G_{1}$, $G_{2}$ and $G_{3}$ are coupling constants, and they have mass dimension $[{\rm Mass}]^{-2}$.
$V$ denotes a real vector multiplet. We consider the gauge degree of freedom as $U(1)_{V}$ for 
the sake of simplicity, and enough for the purpose of this paper.
A chemical potential $\mu$ is introduced by 
$V\equiv 2\mu\bar{\theta}\sigma^{0}\theta$, as a zeroth-component of vector~[9].
Here, the gauge dynamics of $U(1)$ is not considered. 
In this paper, we follow the conventions for metric, gamma matrices, and spinor algebra 
given in the textbook of Wess and Bagger~[19].
The third term might give a dynamically generated Dirac mass,
while a left-handed and a right-handed Majorana masses might be generated
by the fourth and fifth terms, respectively.
We set aside the question on the origin of these nonlinear interaction terms,
and introduce them with an assumption of the existence of some effective attractive interactions in the system. 
The special case $G_{2}=G_{3}$ ( the left-right symmetric ) will be taken
to keep the parity symmetry in our model.
In fact, the spin-singlet Lorentz-scalar BCS pairing gap is given by a parity-invariant combination of 
a left-handed Majorana mass and a right-handed Majorana mass~[17,18].
The original version of the SNJL model will be obtained by $G_{2}=G_{3}=0$.
Thus, we call (1) as a generalized-SNJL model.
By using (1), we examine the dynamical breakings of global $U(1)_{V}$ 
( broken by superconductivity )
and $U(1)_{A}$ ( chiral ) symmetries at finite temperature and density.

Our model Lagrangian (1) will be rewritten in the following form through 
the method of SUSY auxiliary fields~[13,14,15,16]:
\begin{eqnarray}
{\cal L} &=& \int d^{2}\theta d^{2}\bar{\theta}\Bigl( (1-\Delta^{2}\theta^{2}\bar{\theta}^{2})(\Phi^{\dagger}_{+}e^{2\mu\bar{\theta}\sigma^{0}\theta}\Phi_{+}+\Phi^{\dagger}_{-}e^{-2\mu\bar{\theta}\sigma^{0}\theta}\Phi_{-}) + \frac{N}{G_{1}}H^{\dagger}_{1}H_{1} + \frac{N}{G_{2}}H^{\dagger}_{2}H_{2}+ \frac{N}{G_{3}}H^{\dagger}_{3}H_{3} \nonumber \\
& & + \delta(\bar{\theta})S_{1}\bigl( \frac{N}{G_{1}}H_{1}-\Phi_{+}\Phi_{-} \bigr) 
+ \delta(\theta)S^{\dagger}_{1}\bigl( \frac{N}{G_{1}}H^{\dagger}_{1}-\Phi^{\dagger}_{+}\Phi^{\dagger}_{-} \bigr)   \nonumber \\
& & + \delta(\bar{\theta})S_{2}\bigl( \frac{N}{G_{2}}H_{2}-\Phi_{+}\Phi_{+} \bigr) 
+ \delta(\theta)S^{\dagger}_{2}\bigl( \frac{N}{G_{2}}H^{\dagger}_{2}-\Phi^{\dagger}_{+}\Phi^{\dagger}_{+} \bigr)   \nonumber \\
& & + \delta(\bar{\theta})S_{3}\bigl( \frac{N}{G_{3}}H_{3}-\Phi_{-}\Phi_{-} \bigr)  
+ \delta(\theta)S^{\dagger}_{3}\bigl( \frac{N}{G_{3}}H^{\dagger}_{3}-\Phi^{\dagger}_{-}\Phi^{\dagger}_{-} \bigr)  \Bigr). 
\end{eqnarray}
In order to realize the DCSB in this model, we have introduced a soft SUSY breaking mass $\Delta$.
It was shown in the ordinary SNJL model: Because of the nonrenormalization theorem,
the chiral symmetry cannot be broken dynamically if the theory maintains the ${\cal N}=1$ SUSY exactly~[13].
We speculate it is also the case in (1).
Expanding the Lagrangian (2) in terms of the component fields, 
eliminating the auxiliary fields of the chiral multiplets through their Euler equations, 
and keeping only the relevant terms in the leading order in the sense of the large-$N$ expansion, we get
\begin{eqnarray}
{\cal L} &=& -\frac{N}{G_{1}}|\phi_{S_{1}}|^{2}-\frac{N}{G_{2}}|\phi_{S_{2}}|^{2}-\frac{N}{G_{3}}|\phi_{S_{3}}|^{2}  \nonumber \\
& & -(\partial_{\nu}-i\mu\delta_{\nu 0})\phi^{\dagger}_{+}(\partial^{\nu}+i\mu\delta_{\nu 0})\phi_{+}-|\phi_{S_{1}}|^{2}\phi^{\dagger}_{+}\phi_{+}-4|\phi_{S_{2}}|^{2}\phi^{\dagger}_{+}\phi_{+}-\Delta^{2}\phi^{\dagger}_{+}\phi_{+}   \nonumber \\
& & -(\partial_{\nu}+i\mu\delta_{\nu 0})\phi^{\dagger}_{-}(\partial^{\nu}-i\mu\delta_{\nu 0})\phi_{-}-|\phi_{S_{1}}|^{2}\phi^{\dagger}_{-}\phi_{-}-4|\phi_{S_{3}}|^{2}\phi^{\dagger}_{-}\phi_{-}-\Delta^{2}\phi^{\dagger}_{-}\phi_{-}   \nonumber \\
& & -2\phi^{\dagger}_{S_{1}}\phi_{S_{2}}\phi^{\dagger}_{-}\phi_{+}-2\phi^{\dagger}_{S_{2}}\phi_{S_{1}}\phi^{\dagger}_{+}\phi_{-}-2\phi^{\dagger}_{S_{1}}\phi_{S_{3}}\phi^{\dagger}_{+}\phi_{-}-2\phi^{\dagger}_{S_{3}}\phi_{S_{1}}\phi^{\dagger}_{-}\phi_{+}  \nonumber \\
& & -i\bar{\psi}_{+}\bar{\sigma}^{\nu}(\partial_{\nu}-i\mu\delta_{\nu 0})\psi_{+}-i\bar{\psi}_{-}\bar{\sigma}^{\nu}(\partial_{\nu}+i\mu\delta_{\nu 0})\psi_{-}   \nonumber \\
& & +\phi_{S_{1}}\psi_{+}\psi_{-} +\phi^{\dagger}_{S_{1}}\bar{\psi}_{+}\bar{\psi}_{-}+\phi_{S_{2}}\psi_{+}\psi_{+}+\phi^{\dagger}_{S_{2}}\bar{\psi}_{+}\bar{\psi}_{+}+\phi_{S_{3}}\psi_{-}\psi_{-}+\phi^{\dagger}_{S_{3}}\bar{\psi}_{-}\bar{\psi}_{-}.
\end{eqnarray}
Here, $\phi_{S_{j}}$ ( $j=1-3$ ) denote the scalar components of the chiral multiplets $S_{j}$ ( $j=1-3$ ).
We also assumed that, all of $\phi_{S_{j}}$ are constant while all of the spinor components of $S_{j}$ are zero.
This assumption is justified in the leading order of the $1/N$ expansion~[13,14,15,16].
Therefore, the partition function is obtained in the following form:
\begin{eqnarray}
{\cal Z} &=& \int 
{\cal D}\phi_{+}{\cal D}\phi^{\dagger}_{+}{\cal D}\phi_{-}{\cal D}\phi^{\dagger}_{-}{\cal D}\Psi{\cal D}\bar{\Psi}
{\cal D}\phi_{S_{1}}{\cal D}\phi^{\dagger}_{S_{1}}{\cal D}\phi_{S_{2}}{\cal D}\phi^{\dagger}_{S_{2}}{\cal D}\phi_{S_{3}}{\cal D}\phi^{\dagger}_{S_{3}}
\exp\Bigl\{ i\int d^{4}x  \nonumber \\ 
& & \quad \times \Bigl( -\frac{N}{G_{1}}|\phi_{S_{1}}|^{2}-\frac{N}{G_{2}}|\phi_{S_{2}}|^{2}-\frac{N}{G_{3}}|\phi_{S_{3}}|^{2} +\Pi^{\dagger}\Omega_{B}\Pi + \frac{1}{2}\bar{\Xi}\Omega_{F}\Xi \Bigr) \Bigr\}   \nonumber \\
&=& \int {\cal D}\phi_{S_{1}}{\cal D}\phi^{\dagger}_{S_{1}}{\cal D}\phi_{S_{2}}{\cal D}\phi^{\dagger}_{S_{2}}{\cal D}\phi_{S_{3}}{\cal D}\phi^{\dagger}_{S_{3}}
\exp\Bigl\{ iN\int d^{4}x  \nonumber \\ 
& & \quad \times \Bigl( -\frac{1}{G_{1}}|\phi_{S_{1}}|^{2}-\frac{1}{G_{2}}|\phi_{S_{2}}|^{2}-\frac{1}{G_{3}}|\phi_{S_{3}}|^{2} \Bigr) + 2i\ln{\rm Det}\Omega_{B} -i\ln{\rm Det}\Omega_{F}  \Bigr\}, 
\end{eqnarray}
where, we have used the following definitions of the fields:
\begin{eqnarray}
\Pi \equiv 
\left( 
\begin{array}{c}
\phi_{+} \\
\phi_{-} 
\end{array} 
\right), 
\quad 
\Psi \equiv 
\left( 
\begin{array}{c}
\psi_{+} \\
\bar{\psi}_{-} 
\end{array} 
\right),
\quad
\Xi \equiv 
\left(
\begin{array}{c}
\Psi \\
\bar{\Psi}^{T}
\end{array}
\right), 
\quad 
\bar{\Xi} \equiv ( \bar{\Psi}, \Psi^{T} ).   
\end{eqnarray}
Here, $\Psi$ is the four-component Dirac bispinor, 
$\Xi$ and $\bar{\Xi}$ are the eight-component Nambu notations~[17,18,20], and $T$ means the transposition.
The definitions of the matrices $\Omega_{B}$ and $\Omega_{F}$ are given as follows: 
\begin{eqnarray}
\Omega_{B} &\equiv& \left(
\begin{array}{cccc}
(\partial_{\nu}+i\mu\delta_{\nu 0})(\partial^{\nu}+i\mu\delta_{\nu 0})-|\phi_{S_{1}}|^{2}-4|\phi_{S_{2}}|^{2}-\Delta^{2} & -2(\phi^{\dagger}_{S_{2}}\phi_{S_{1}} +\phi^{\dagger}_{S_{1}}\phi_{S_{3}} )  \\
-2(\phi^{\dagger}_{S_{1}}\phi_{S_{2}} +\phi^{\dagger}_{S_{3}}\phi_{S_{1}} )  &  (\partial_{\nu}-i\mu\delta_{\nu 0})(\partial^{\nu}-i\mu\delta_{\nu 0})-|\phi_{S_{1}}|^{2}-4|\phi_{S_{3}}|^{2}-\Delta^{2}
\end{array}
\right),  \\
\Omega_{F} &\equiv& \left(
\begin{array}{cccc}
i\partfey+\gamma^{0}\mu-\frac{1}{2}\phi_{S_{1}}(1+i\gamma_{5})-\frac{1}{2}\phi^{\dagger}_{S_{1}}(1-i\gamma_{5}) & \phi_{S_{3}}C(1+i\gamma_{5})-\phi^{\dagger}_{S_{2}}C(1-i\gamma_{5}) \\
-\phi_{S_{2}}C(1+i\gamma_{5})+\phi^{\dagger}_{S_{3}}C(1-i\gamma_{5}) & i\partfey^{T}-\gamma^{0T}\mu+\frac{1}{2}\phi_{S_{1}}(1+i\gamma_{5})+\frac{1}{2}\phi^{\dagger}_{S_{1}}(1-i\gamma_{5})
\end{array}
\right), 
\end{eqnarray}
where, $\gamma_{5}\equiv\gamma^{0}\gamma^{1}\gamma^{2}\gamma^{3}$, 
and $C\equiv i\gamma^{2}\gamma^{0}$ is the charge conjugation matrix.
It is a well-known fact that, in the relativistic theory of superconductivity,
we should take the parity-invariant Lorentz-scalar symmetry for the spin singlet BCS pairing state~[6,17,18].
In order to take the symmetries of both the Dirac mass and the BCS pairing gap in $\Omega_{F}$ as Lorentz scalar,
we choose $\phi^{\dagger}_{S_{1}}=\phi_{S_{1}}$ and $\phi^{\dagger}_{S_{2}}=\phi_{S_{3}}$.
At $\phi^{\dagger}_{S_{2}}\ne\phi_{S_{3}}$, the gap function corresponds to a
linear combination of the scalar and pseudoscalar pairings.
The pseudoscalar pairing cannot realize in the relativistic BCS theory~[17],
and thus we may introduce the constraint $\phi^{\dagger}_{S_{2}}=\phi_{S_{3}}$ also in the present case. 
The boson and fermion determinants in Eq. (4) will be evaluated in the following forms:
\begin{eqnarray}
{\rm det}\Omega_{B} &=& (p_{0}-E^{B}_{+}(\bmp))(p_{0}+E^{B}_{+}(\bmp))(p_{0}-E^{B}_{-}(\bmp))(p_{0}+E^{B}_{-}(\bmp)),  \\
{\rm det}\Omega_{F} &=& (p_{0}-E^{F}_{+}(\bmp))^{2}(p_{0}+E^{F}_{+}(\bmp))^{2}(p_{0}-E^{F}_{-}(\bmp))^{2}(p_{0}+E^{F}_{-}(\bmp))^{2}.  
\end{eqnarray}
All of the eigenvalues of $\Omega_{F}$ doubly degenerate.
These degeneracies relate to the time-reversal invariance of the BCS gap function~[17].
The bosonic and fermionic quasiparticle energy spectra are evaluated as follows:
\begin{eqnarray}
E^{B}_{+}(\bmp) &\equiv& \sqrt{ \bmp^{2}+\mu^{2}+|\phi_{S_{1}}|^{2}+4|\phi_{S_{2}}|^{2}+\Delta^{2}-2\sqrt{ \mu^{2}(\bmp^{2}+|\phi_{S_{1}}|^{2}+4|\phi_{S_{2}}|^{2}+\Delta^{2})+4|\phi_{S_{1}}|^{2}|\phi_{S_{2}}|^{2}   } },  \\
E^{B}_{-}(\bmp) &\equiv& \sqrt{ \bmp^{2}+\mu^{2}+|\phi_{S_{1}}|^{2}+4|\phi_{S_{2}}|^{2}+\Delta^{2}+2\sqrt{ \mu^{2}(\bmp^{2}+|\phi_{S_{1}}|^{2}+4|\phi_{S_{2}}|^{2}+\Delta^{2})+4|\phi_{S_{1}}|^{2}|\phi_{S_{2}}|^{2}   } },  \\
E^{F}_{+}(\bmp) &\equiv& \sqrt{(\sqrt{\bmp^{2}+|\phi_{S_{1}}|^{2}}-\mu)^{2}+4|\phi_{S_{2}}|^{2}},  \\
E^{F}_{-}(\bmp) &\equiv& \sqrt{(\sqrt{\bmp^{2}+|\phi_{S_{1}}|^{2}}+\mu)^{2}+4|\phi_{S_{2}}|^{2}}.
\end{eqnarray}
We confirm $|\phi_{S_{1}}|$ is the dynamically generated Dirac mass, while $2|\phi_{S_{2}}|$ 
corresponds to the BCS gap function.

As we mentioned above, we consider the case $G_{2}=G_{3}$.
The effective action of the leading order in the large-$N$ expansion is found to be
\begin{eqnarray}
\Gamma_{eff} &=& \int d^{4}x\bigl( -\frac{|\phi_{S_{1}}|^{2}}{G_{1}} -2\frac{|\phi_{S_{2}}|^{2}}{G_{2}}  \bigr) +2i\ln{\rm Det}\Omega_{B} -i\ln{\rm Det}\Omega_{F}. 
\end{eqnarray}
Hereafter, we introduce the finite-temperature Matsubara formalism~[21].
The Matsubara formalism is obtained by the following substitutions in our theory:
\begin{eqnarray}
\int\frac{dp_{0}}{2\pi i} \to \sum_{n}\frac{1}{\beta}, \qquad p_{0} \to i\omega^{B}_{n}, i\omega^{F}_{n},
\end{eqnarray}
where $\beta\equiv 1/k_{B}T$ ( $k_{B}$; the Boltzmann constant, $T$; temperature ). 
$k_{B}=1$ is taken throughout this paper.
$\omega^{B}_{n}$ and $\omega^{F}_{n}$ are the boson and fermion discrete frequencies, respectively.
Their definitions are $\omega^{B}_{n}\equiv 2n\pi/\beta$ and $\omega^{F}_{n}\equiv (2n+1)\pi/\beta$ 
( $n=0,\pm 1,\pm 2, \cdots$ ). 
The effective potential ( thermodynamic potential ) becomes
\begin{eqnarray}
V_{eff}(\phi_{S_{1}},\phi_{S_{2}}) &=& \frac{|\phi_{S_{1}}|^{2}}{G_{1}} + 2\frac{|\phi_{S_{2}}|^{2}}{G_{2}} + \sum_{n}\frac{1}{\beta}\int^{\Lambda}\frac{d^{3}\bmp}{(2\pi)^{3}}\ln\det\Omega_{F} -2\sum_{n}\frac{1}{\beta}\int^{\Lambda}\frac{d^{3}\bmp}{(2\pi)^{3}}\ln\det\Omega_{B}   \nonumber  \\
&=& \frac{|\phi_{S_{1}}|^{2}}{G_{1}} + 2\frac{|\phi_{S_{2}}|^{2}}{G_{2}} + \int^{\Lambda}\frac{d^{3}\bmp}{(2\pi)^{3}}\Bigg( E^{B}_{+}(\bmp)+E^{B}_{-}(\bmp)-E^{F}_{+}(\bmp)-E^{F}_{-}(\bmp)    \nonumber \\ 
& & \qquad -\frac{2}{\beta}\ln(1+e^{-\beta E^{F}_{+}(\bmp)})(1+e^{-\beta E^{F}_{-}(\bmp)})+\frac{2}{\beta}\ln(1-e^{-\beta E^{B}_{+}(\bmp)})(1-e^{-\beta E^{B}_{-}(\bmp)})  \Bigg). 
\end{eqnarray}
To obtain the final expression in Eq. (16), the frequency summations were performed. 
A three-dimensional momentum cutoff $\Lambda$ was introduced to regularize the integral.
The gap equations are derived in the following forms: 
\begin{eqnarray}
0 &=& \frac{\partial V_{eff}}{\partial |\phi_{S_{1}}| }  \nonumber \\
&=& \frac{2|\phi_{S_{1}}|}{G_{1}} - |\phi_{S_{1}}|\int^{\Lambda}\frac{d^{3}\bmp}{(2\pi)^{3}}\Bigl\{ \bigl( 1-\frac{\mu}{\sqrt{\bmp^{2}+|\phi_{S_{1}}|^{2} } }\bigr)\frac{1}{E^{F}_{+}}\tanh\frac{\beta}{2}E^{F}_{+}+\bigl( 1+\frac{\mu}{\sqrt{\bmp^{2}+|\phi_{S_{1}}|^{2} } }\bigr)\frac{1}{E^{F}_{-}}\tanh\frac{\beta}{2}E^{F}_{-} \Bigr\}  \nonumber \\
& & + |\phi_{S_{1}}|\int^{\Lambda}\frac{d^{3}\bmp}{(2\pi)^{3}}\Bigl\{ \bigl( 1-\frac{\mu^{2}+4|\phi_{S_{2}}|^{2}}{\sqrt{\mu^{2}(\bmp^{2}+|\phi_{S_{1}}|^{2}+4|\phi_{S_{2}}|^{2}+\Delta^{2} )+4|\phi_{S_{1}}|^{2}|\phi_{S_{2}}|^{2}}} \bigr)\frac{1}{E^{B}_{+}}\coth\frac{\beta}{2}E^{B}_{+}   \nonumber \\
& & \quad + \bigl( 1+\frac{\mu^{2}+4|\phi_{S_{2}}|^{2}}{\sqrt{\mu^{2}(\bmp^{2}+|\phi_{S_{1}}|^{2}+4|\phi_{S_{2}}|^{2}+\Delta^{2} )+4|\phi_{S_{1}}|^{2}|\phi_{S_{2}}|^{2}}} \bigr)\frac{1}{E^{B}_{-}}\coth\frac{\beta}{2}E^{B}_{-}  \Bigr\},    \\
0 &=& \frac{\partial V_{eff}}{\partial |\phi_{S_{2}}| }  \nonumber \\
&=& \frac{4|\phi_{S_{2}}|}{G_{2}} - 4|\phi_{S_{2}}|\int^{\Lambda}\frac{d^{3}\bmp}{(2\pi)^{3}}\Bigl\{ \frac{1}{E^{F}_{+}}\tanh\frac{\beta}{2}E^{F}_{+}+\frac{1}{E^{F}_{-}}\tanh\frac{\beta}{2}E^{F}_{-} \Bigr\}  \nonumber \\
& & + 4|\phi_{S_{2}}|\int^{\Lambda}\frac{d^{3}\bmp}{(2\pi)^{3}}\Bigl\{ \bigl( 1-\frac{\mu^{2}+|\phi_{S_{1}}|^{2}}{\sqrt{\mu^{2}(\bmp^{2}+|\phi_{S_{1}}|^{2}+4|\phi_{S_{2}}|^{2}+\Delta^{2} )+4|\phi_{S_{1}}|^{2}|\phi_{S_{2}}|^{2}}} \bigr)\frac{1}{E^{B}_{+}}\coth\frac{\beta}{2}E^{B}_{+}   \nonumber \\
& & \quad + \bigl( 1+\frac{\mu^{2}+|\phi_{S_{1}}|^{2}}{\sqrt{\mu^{2}(\bmp^{2}+|\phi_{S_{1}}|^{2}+4|\phi_{S_{2}}|^{2}+\Delta^{2} )+4|\phi_{S_{1}}|^{2}|\phi_{S_{2}}|^{2}}} \bigr)\frac{1}{E^{B}_{-}}\coth\frac{\beta}{2}E^{B}_{-}  \Bigr\}.   
\end{eqnarray}
At first glance, both of these gap equations seem to include quadratic divergences, and they might be influenced
by $\Delta$. The effect of SUSY is parametrized by $\Delta$.
These gap equations correctly give their limiting cases at $\Delta\to \infty$.
For example, Eq. (18) gives the gap equation of the non-SUSY relativistic superconductivity~[17] 
at $\Delta\to \infty$.
Equations (17) and (18) can have nontrivial solutions at least at $G_{1},G_{2}>0$ ( attractive interactions ).
The charge density $\varrho$, the conjugate of $\mu$, is found to be
\begin{eqnarray}
\varrho &=& -\frac{\partial V_{eff}}{\partial \mu} \nonumber \\
&=& \int\frac{d^{3}\bmp}{(2\pi)^{3}} \Bigg( 
\frac{\mu-\sqrt{\bmp^{2}+|\phi_{S_{1}}|^{2}}}{E^{F}_{+}}\tanh\frac{\beta}{2}E^{F}_{+}+\frac{\mu+\sqrt{\bmp^{2}+|\phi_{S_{1}}|^{2}}}{E^{F}_{-}}\tanh\frac{\beta}{2}E^{F}_{-}  \nonumber \\
& & \quad -\Bigl( 1-\frac{\bmp^{2}+|\phi_{S_{1}}|^{2}+4|\phi_{S_{2}}|^{2}+\Delta^{2}}{\sqrt{\mu^{2}(\bmp^{2}+|\phi_{S_{1}}|^{2}+4|\phi_{S_{2}}|^{2}+\Delta^{2} )+4|\phi_{S_{1}}|^{2}|\phi_{S_{2}}|^{2}}} \Bigr)\frac{\mu}{E^{B}_{+}}\coth\frac{\beta}{2}E^{B}_{+}   \nonumber \\
& & \quad -\Bigl( 1+\frac{\bmp^{2}+|\phi_{S_{1}}|^{2}+4|\phi_{S_{2}}|^{2}+\Delta^{2}}{\sqrt{\mu^{2}(\bmp^{2}+|\phi_{S_{1}}|^{2}+4|\phi_{S_{2}}|^{2}+\Delta^{2} )+4|\phi_{S_{1}}|^{2}|\phi_{S_{2}}|^{2}}} \Bigr)\frac{\mu}{E^{B}_{-}}\coth\frac{\beta}{2}E^{B}_{-}   \Bigg).   
\end{eqnarray}
In the zero-temperature case with $|\phi_{S_{2}}|=0$, one has
\begin{eqnarray}
\varrho &=& 2\int \frac{d^{3}\bmp}{(2\pi)^{3}} \theta(\mu-\sqrt{\bmp^{2}+|\phi_{S_{1}}|^{2}}) = \frac{p^{3}_{F}}{3\pi^{2}}, 
\end{eqnarray}
where, $\theta(x)$ is the step function defined as follows: $\theta(x)=1$ for $x>0$ and $\theta(x)=0$ for $x<0$.
$p_{F}$ is the Fermi momentum.
At $T=0$, $\mu$ coincides with the Fermi energy $\sqrt{p^{2}_{F}+|\phi_{S_{1}}|^{2}}$,
and it is determined by the charge density of fermion. 
Later, Eq. (18) is solved numerically.
We completely neglect the temperature dependence of $\mu$, 
and use it as an external parameter for our numerical calculations.

\section{Quasiparticle Excitation Energy Spectra}

In this section, we examine the quasiparticle excitation energy spectra (10)-(13).
Because we consider several situations of $\mu\ne 0$, the Bose-Einstein condensation ( BEC ) can 
take place in our model~[9,21]. 
Depending on the model parameters ( $\Delta$, $\mu$, etc. ) and $|\phi_{S_{1}}|$ and $|\phi_{S_{2}}|$,
$E^{B}_{+}(\bmp)$ can have a zero-point. 
In such a case, the logarithmic function $\ln (1-e^{-\beta E^{B}_{+}})$ in $V_{eff}$ 
will diverge and BEC takes place. A similar discussion on BEC was given in Ref.~[9].
In this papar, we will not study the physical property of the BEC phase in our model,
and only discuss the phase boundary in the model-parameter space.
This can be done by the examination of the quasiparticle energy spectra.
We have to find the condition of the realization of BEC before solving the gap equations (17) and (18) 
to study the phases of the DCSB and the superconductivity.

First, we examine Eqs. (10)-(13) in several limiting cases.
At $\mu=0$, the zero-density case, the energy spectra becomes
\begin{eqnarray}
E^{B}_{\pm}(\bmp) &=& \sqrt{\bmp^{2}+(|\phi_{S_{1}}|\mp2|\phi_{S_{2}}|)^{2}+\Delta^{2}},   \\
E^{F}_{\pm}(\bmp) &=& \sqrt{\bmp^{2}+|\phi_{S_{1}}|^{2}+4|\phi_{S_{2}}|^{2}}.
\end{eqnarray}
On the other hand, at $|\phi_{S_{2}}|=0$ one finds 
the spectra of the DCSB at finite density:
\begin{eqnarray}
E^{B}_{\pm}(\bmp) &=& \sqrt{\bmp^{2}+|\phi_{S_{1}}|^{2}+\Delta^{2}}\mp\mu, \\
E^{F}_{\pm}(\bmp) &=& \sqrt{\bmp^{2}+|\phi_{S_{1}}|^{2}}\mp\mu,
\end{eqnarray}
while, at $|\phi_{S_{1}}|=0$ where only a superconducting gap 
is generated, the spectra take the following forms:
\begin{eqnarray}
E^{B}_{\pm}(\bmp) &=& \sqrt{\bmp^{2}+4|\phi_{S_{2}}|^{2}+\Delta^{2}}\mp\mu, \\
E^{F}_{\pm}(\bmp) &=& \sqrt{(|\bmp|\mp\mu)^{2}+4|\phi_{S_{2}}|^{2}}.
\end{eqnarray}
In Eqs. (23) and (25), $E^{B}_{+}(\bmp)$ can become negative with $\mu\ge 0$.
Because of the positiveness of the Bose distribution function $1/(e^{\beta E^{B}_{+}}-1)$,
and $\mu$ corresponds to the Fermi energy of the system at zero temperature, 
\begin{eqnarray}
|\phi_{S_{1}}|^{2} \le \mu^{2} \le |\phi_{S_{1}}|^{2} + \Delta^{2}
\end{eqnarray}
has to be satisfied in Eqs. (23) and (24). 
If we use $\mu=\sqrt{p^{2}_{F}+|\phi_{S_{1}}|^{2}}$, (27) will be rewritten as
the condition of the external parameters $p_{F}$ and $\Delta$:
\begin{eqnarray}
0 \le p_{F} \le \Delta.
\end{eqnarray} 
Thus, $\Delta$ is the upperbound for $p_{F}$ in the SUSY theory.
For Eqs. (25) and (26),
\begin{eqnarray}
0 \le \mu^{2}=p^{2}_{F} \le 4|\phi_{S_{2}}|^{2}+\Delta^{2}
\end{eqnarray}
has to be satisfied. 
From (23) one finds that,
if dynamically generated $|\phi_{S_{1}}|$ obtained as a solution of the gap equation (17)
satisfies $\mu^{2}=|\phi_{S_{1}}|^{2} + \Delta^{2}$, 
the BEC takes place at the mode of $\bmp=0$.

Figure 1 shows a typical case of the excitation energy spectra of boson and fermion quasiparticles
under the superconducting state with a non-vanishing chiral mass.
$E^{F}_{+}(\bmp)$ has a minimum at $p_{F}=\sqrt{\mu^{2}-|\phi_{S_{1}}|^{2}}$ 
where the excitation energy gap $4|\phi_{S_{2}}|$ locates.
The appearance of the branches $E^{B}_{+}(\bmp)$ and $E^{B}_{-}(\bmp)$ is the new phenomenon
of our SUSY theory compared with the no-SUSY relativistic BCS superconductivity~[6,17,18].
The Bose branch $E^{B}_{+}$ has an energy gap at $|\bmp|=0$.
All of the spectra become parallel with the light cone at $|\bmp|\to\infty$.
Let us examine the situation $E^{B}_{+}(\bmp)=0$ in detail.
From Eq. (10), one obtains the solutions of the equation $E^{B}_{+}(\bmp)=0$ as 
\begin{eqnarray}
|\bmp| &=& \sqrt{\mu^{2}-|\phi_{S_{1}}|^{2}-4|\phi_{S_{2}}|^{2}-\Delta^{2}\pm 4|\phi_{S_{1}}||\phi_{S_{2}}|}.
\end{eqnarray}
From the positiveness of the Bose distribution function, 
we have to avoid the situation where $E^{B}_{+}(\bmp)=0$ has two solutions,
because any $|\bmp|$ inside these two solutions makes $E^{B}_{+}(\bmp)$ a complex number.
Therefore, we consider the following three cases in Eq. (30): 
(I) $|\phi_{S_{1}}|=0$ and $|\phi_{S_{2}}|\ne0$ ( superconducting state without DCSB ), 
(II) $|\phi_{S_{2}}|=0$ and $|\phi_{S_{1}}|\ne0$ ( DCSB without superconductivity ),
(III) $|\phi_{S_{1}}|=|\phi_{S_{2}}|=0$.
In the case (I), Eq. (30) becomes $|\bmp|=\sqrt{\mu^{2}-4|\phi_{S_{2}}|^{2}-\Delta^{2}}$.
From this solution, one finds the condition $\mu^{2}\ge 4|\phi_{S_{2}}|^{2}+\Delta^{2}$ has to be satisfied.
Thus, taking into account (29), we get the critical SUSY soft mass of the superconductivity:
\begin{eqnarray}
\Delta^{sc (I)}_{cr} &=& \sqrt{\mu^{2}-4|\phi_{S_{2}}|^{2}} \sim p_{F}.
\end{eqnarray} 
Here, we used $\mu =p_{F}$.
When $\Delta=\Delta^{sc (I)}_{cr}$ in the case (I), 
$E^{B}_{+}(\bmp)$ has one zero-point at $|\bmp|=0$, and it behaves at $|\bmp|\to 0$ as 
$E^{B}_{+}(\bmp) \approx \bmp^{2}/2\sqrt{(\Delta^{sc(I)}_{cr})^{2}+4|\phi_{S_{2}}|^{2}}$.
Thus we conclude that, in the case (I), 
the superconductivity can coexist with the BEC at $\Delta=\Delta^{sc (I)}_{cr}$,
while only the superconductivity can take place at $\Delta>\Delta^{sc (I)}_{cr}$.
In other words, the phase boundary given by (31) depends on a self-consistently determined $|\phi_{S_{2}}|$.
In the case (II), we find $|\bmp|=\sqrt{\mu^{2}-|\phi_{S_{1}}|^{2}-\Delta^{2}}$ 
as one solution of $E^{B}_{+}(\bmp)=0$.
From this solution, one has $\mu^{2}\ge |\phi_{S_{1}}|^{2}+\Delta^{2}$,
and with the condition (27), we obtain the critical SUSY soft mass of the superconductivity:
\begin{eqnarray}
\Delta^{sc (II)}_{cr} &=& \sqrt{\mu^{2}-|\phi_{S_{1}}|^{2}} = p_{F}.
\end{eqnarray}
Here, we used $\mu =\sqrt{p^{2}_{F}+|\phi_{S_{1}}|^{2}}$.
When we choose the model parameters suitably, especially to satisfy $\Delta > \Delta^{sc (II)}_{cr}$ 
with $|\phi_{S_{2}}|\ne 0$,
$E^{B}_{+}(\bmp)$ takes a positive value for any $|\bmp|$ as shown in Fig. 1. 
On the other hand, the BEC in the boson sector takes place at $\Delta=\Delta^{sc(II)}_{cr}$, 
and $E^{B}_{+}(\bmp)$ behaves as 
$E^{B}_{+}(\bmp) \approx \bmp^{2}/2\sqrt{(\Delta^{sc(II)}_{cr})^{2}+|\phi_{S_{1}}|^{2}}$ at $|\bmp|\to 0$ 
( there is no energy gap of the branch $E^{B}_{+}$ ).
In this situation, $|\phi_{S_{2}}|=0$ has to be satisfied, thus the superconductivity cannot coexist with the BEC.
We find that the chiral symmetry ( chiral mass $|\phi_{S_{1}}|$ ) affects the physical situation of 
the phase boundary between the superconductivity and the BEC. These results are summarized in table I.

\section{Dynamical Chiral Symmetry Breaking}

Now, we arrive at the stage to study the DCSB without superconductivity.
At $T=0$ and $|\phi_{S_{2}}|=0$, the gap equation (17) becomes
\begin{eqnarray}
1 &=& \frac{G_{1}}{2\pi^{2}}\int^{\Lambda}_{0}p^{2}dp\Bigl( \frac{\theta(p-p_{F})}{\sqrt{p^{2}+|\phi_{S_{1}}|^{2}}} -\frac{1}{\sqrt{p^{2}+|\phi_{S_{1}}|^{2}+\Delta^{2}}} \Bigr)  \nonumber \\
&=& \frac{G_{1}}{4\pi^{2}}\Bigl( \Lambda\sqrt{\Lambda^{2}+|\phi_{S_{1}}|^{2}}-|\phi_{S_{1}}|^{2}\ln\frac{\Lambda+\sqrt{\Lambda^{2}+|\phi_{S_{1}}|^{2}}}{|\phi_{S_{1}}|}   \nonumber \\
& & \qquad -\Lambda\sqrt{\Lambda^{2}+|\phi_{S_{1}}|^{2}+\Delta^{2}} + (|\phi_{S_{1}}|^{2}+\Delta^{2})\ln\frac{\Lambda+\sqrt{\Lambda^{2}+|\phi_{S_{1}}|^{2}+\Delta^{2}}}{\sqrt{|\phi_{S_{1}}|^{2}+\Delta^{2}}} \nonumber \\
& & \qquad -p_{F}\sqrt{p^{2}_{F}+|\phi_{S_{1}}|^{2}}+|\phi_{S_{1}}|^{2}\ln\frac{p_{F}+\sqrt{p^{2}_{F}+|\phi_{S_{1}}|^{2}}}{|\phi_{S_{1}}|}   \Bigr).
\end{eqnarray} 
With taking the limit $\Delta\to \infty$, 
Eq. (33) gives the gap equation of the DCSB
of the ordinary ( no-SUSY ) Nambu$-$Jona-Lasinio model at finite density~[5], 
while at $p_{F}=0$ it will give the gap equation of the zero-density case of the SNJL model~[13,15].
This equation determines the phase diagram in the space of parameters 
$G_{1}\Lambda^{2}$, $\Delta/\Lambda$ and $p_{F}/\Lambda$.
The examination of the critical coupling might be the easiest method
to see the relation of several possible phases.
The determination equation for the critical coupling $(G_{1})_{cr}$ is obtained as follows:
\begin{eqnarray}
(G_{1})_{cr}\Lambda^{2} &=& \frac{4\pi^{2}}{1-\sqrt{1+\frac{\Delta^{2}}{\Lambda^{2}}}+\frac{\Delta^{2}}{\Lambda^{2}}\ln\frac{1+\sqrt{1+\Delta^{2}/\Lambda^{2}}}{\Delta/\Lambda}-\frac{p^{2}_{F}}{\Lambda^{2}}}.
\end{eqnarray}
Figure 2 shows $(G_{1})_{cr}\Lambda^{2}$ as a function of $\Delta/\Lambda$.
Equation (34) gives $(G_{1})_{cr}\Lambda^{2}$ of the no-SUSY case as 
$\lim_{\Delta/\Lambda\to\infty}(G_{1})_{cr}\Lambda^{2}=4\pi^{2}/(1-p^{2}_{F}/\Lambda^{2})$.
The denominator of Eq. (34) includes the finite-density effect on $(G_{1})_{cr}$.
$(G_{1})_{cr}$ of $p_{F}\ne 0$ is larger than that of $p_{F}=0$.
The divergence of $(G_{1})_{cr}\Lambda^{2}$ at 
a non-zero value of $\Delta/\Lambda$ ( depends on a numerical value of $p_{F}$ )
indicates the existence of the critical soft mass $\Delta^{dcsb}_{cr}$ for the DCSB of this model~[13,15]. 
From (28), we know $p_{F}\le\Delta$ has to be satisfied in Fig. 2:
When $p_{F}=\Delta$, the BEC takes place at the mode of $\bmp=0$.

\section{Superconductivity}

Now, we consider the following situation:
After the Dirac mass $|\phi_{S_{1}}|$ dynamically generated and the chiral symmetry was broken,
we assume the superconducting instability occurs in the system. 
We solve the gap equation (18) under this situation, 
and restrict ourselves to examining the superconductivity.
Thus, $|\phi_{S_{1}}|$ is treated as a model parameter.
Usually in various superconductors we know, 
the BCS gap function $2|\phi_{S_{2}}|$ is much smaller than $\mu-|\phi_{S_{1}}|$~[18]. 
We assume it is also the case in our numerical calculation for solving Eq. (18),
and choose the model parameters $G_{2}$, $\Lambda$, $\mu$, $\Delta$ and $|\phi_{S_{1}}|$ 
to satisfy $2|\phi_{S_{2}}|/(\mu-|\phi_{S_{1}}|)\ll 1$.
Because $2|\phi_{S_{2}}|$ is quite sensitive to $G_{2}$ and $\Lambda$ even in our SUSY theory, 
we carefully choose numerical values of them 
( a kind of fine-tuning ) to get a physically reasonable solution.
In this paper, we set aside the question on the origin of the interactions given from the nonlinear terms
of $\Phi_{\pm}$. We treat the cutoff in a general way. 
As discussed in Sec.III, the superconductivity never occurs at $p_{F}=\Delta$ with $|\phi_{S_{1}}|\ne 0$.
We also have to take into account the condition (28) to choose a numerical value of $\Delta$;
when $\Delta\le\Delta^{sc(II)}_{cr}=p_{F}$ is satisfied, the superconductivity cannot realize.

Figure 3 shows the gap function $2|\phi_{S_{2}}|$ at $T=0$ as a function of the coupling constant $G_{2}$.
Equation (18) was solved under the three examples: $\Delta=2$, $\Delta=10$ and $\Delta=100$
with the energy unit $|\phi_{S_{1}}|=1$. 
The result of the no-SUSY case 
( obtained by dropping the contribution coming from the bosonic part in Eq.(18) ), 
namely the ordinary relativistic superconductivity, is also given in this figure.
All of the examples show qualitatively similar dependences on $G_{2}$,
and they reflect the nonperturbative effect.
When the soft mass becomes large, the dependence on $G_{2}$ becomes strong,
however, the dependence is qualitatively unchanged whether a numerical value of $\Delta$ 
is larger or smaller than $\Lambda$.
The result in Fig. 3 also indicates the absence of the critical coupling 
of the superconductivity similar to the ordinary BCS theory~[1,6,17,18]:
Any attractive interaction gives a superconducting instability also in the SUSY case.
In Eq.(18), the contribution coming from the vicinity of the Fermi energy
dominantly determines the solution $|\phi_{S_{2}}|$.
This contribution can be made arbitrarily large by changing $|\phi_{S_{2}}|$. 
The integrals given from the momentum-integration of the functions 
of the branches $E^{F}_{-}$, $E^{B}_{+}$ and $E^{B}_{-}$ are almost constants 
under the variation of $|\phi_{S_{2}}|$. 
Thus, Eq. (18) at $T=0$ always has one nontrivial solution with any attractive interaction. 
We also find that, Eq. (18) never has two solutions in our numerical calculations. 
In other words, the effective potential (16)
only has one minimum with respect to the variation of $|\phi_{S_{2}}|$
under the superconducting phase.
We confirmed this fact by our numerical calculation of the effective potential.

In Fig. 4, we show the gap function $2|\phi_{S_{2}}|$ at $T=0$ as a function of 
the three-dimensional momentum cutoff $\Lambda$.
In the case $\Delta=2$, the divergence of the gap function 
becomes slower than other examples because of the effect of SUSY.
In this figure, the gap function at $\Delta=2$ diverges almost linearly.
In the cases of $\Delta=10$, $\Delta=100$ and no-SUSY, the gap function diverges almost quadratically 
at $\Lambda\to \infty$.

Figure 5 gives $2|\phi_{S_{2}}|$ as a function of temperature $T$.
$2|\phi_{S_{2}}|$ continuously vanishes at $T\to T_{c}$ ( $T_{c}$; the critical temperature ),
clearly shows the character of second-order phase transition.
In all of the examples shown in this figure, the BCS universal constant
$2|\phi_{S_{2}}(T=0)|/T_{c}=1.76$ is satisfied~[1,18]. 
This fact means that, both $2|\phi_{S_{2}}(T=0)|$ and $T_{c}$ depend on $G_{2}$, $\Lambda$, $\Delta$ and $\mu$,
while their ratio is independent on them.
In the non-relativistic and relativistic BCS theories,
any spin-singlet ( or, Lorentz-scalar ) BCS gap function satisfies the universal constant~[1,18].
We find it is also satisfied in our SUSY BCS theory.

We discuss the thermodynamic property of the SUSY superconductivity.
With taking into account $\partial V_{eff}/\partial |\phi_{S_{2}}|=0$, the entropy $S$ is obtained as follows:
\begin{eqnarray}
S &=& -\frac{\partial V_{eff}}{\partial T} \nonumber \\
&=& 2\int^{\Lambda}\frac{d^{3}\bmp}{(2\pi)^{3}}\Bigl\{ \ln\frac{(1+e^{-\beta E^{F}_{+}})(1+e^{-\beta E^{F}_{-}})}{(1-e^{-\beta E^{B}_{+}})(1-e^{-\beta E^{B}_{-}})}+\frac{\beta E^{F}_{+}}{e^{\beta E^{F}_{+}}+1} +\frac{\beta E^{F}_{-}}{e^{\beta E^{F}_{-}}+1} +\frac{\beta E^{B}_{+}}{e^{\beta E^{B}_{+}}-1} +\frac{\beta E^{B}_{-}}{e^{\beta E^{B}_{-}}-1}   \Bigr\}. 
\end{eqnarray} 
This is the entropy of the ideal gas of quasiparticles of 
the branches $E^{F}_{+}$, $E^{F}_{-}$, $E^{B}_{+}$ and $E^{B}_{-}$.
At $|\phi_{S_{2}}|\to 0$, (35) gives the entropy of the normal state.
By the result shown in Fig. 5, we find $S$ is continuous at the phase transition:
It is a second-order phase transition. The heat capacity $C$ becomes
\begin{eqnarray}
C &=& T\frac{\partial S}{\partial T} \nonumber \\
&=& 2\int^{\Lambda}\frac{d^{3}\bmp}{(2\pi)^{3}}\Bigg\{ \Bigg( \Bigl(\frac{E^{F}_{+}}{T}\Bigr)^{2}-\frac{E^{F}_{+}}{T}\frac{\partial E^{F}_{+}}{\partial T}  \Bigg)\frac{1}{e^{E^{F}_{+}/T}+1}\Bigl(1-\frac{1}{e^{E^{F}_{+}/T}+1} \Bigr)  \nonumber \\
& & \qquad + \Bigg( \Bigl(\frac{E^{F}_{-}}{T}\Bigr)^{2}-\frac{E^{F}_{-}}{T}\frac{\partial E^{F}_{-}}{\partial T}  \Bigg)\frac{1}{e^{E^{F}_{-}/T}+1}\Bigl(1-\frac{1}{e^{E^{F}_{-}/T}+1} \Bigr) \nonumber \\
& & \qquad + \Bigg( \Bigl(\frac{E^{B}_{+}}{T}\Bigr)^{2}-\frac{E^{B}_{+}}{T}\frac{\partial E^{B}_{+}}{\partial T}  \Bigg)\frac{1}{e^{E^{B}_{+}/T}-1}\Bigl(1+\frac{1}{e^{E^{B}_{+}/T}-1} \Bigr) \nonumber \\
& & \qquad + \Bigg( \Bigl(\frac{E^{B}_{-}}{T}\Bigr)^{2}-\frac{E^{B}_{-}}{T}\frac{\partial E^{B}_{-}}{\partial T}  \Bigg)\frac{1}{e^{E^{B}_{-}/T}-1}\Bigl(1+\frac{1}{e^{E^{B}_{-}/T}-1} \Bigr)\Bigg\}.
\end{eqnarray}
Let us recall the energy spectra shown in Fig. 1.
Because the density of states diverges around the energy gap of $E^{F}_{+}$~[18], 
the contribution of the excitation of $E^{F}_{+}$ dominates the integrals of $S$ and $C$, 
and other branches give almost no contribution. Both the entropy $S$ and the heat capacity $C$ 
of the system are determined by the thermal excitations of the quasiparticles of the branch $E^{F}_{+}$. 
Because the temperature dependence of $2|\phi_{S_{2}}|$ is the same with the well-known BCS result,
we conclude that the temperature dependences of $S$ and $C$ are qualitatively the same 
with the well-known results of the ordinary BCS theory~[1,18]:
The heat capacity $C$ of Eq. (36) becomes exponentially small in the limit $T\to 0$, 
while it behaves as $C\propto T$ ( the Fermi liquid behavior ) at $T>T_{c}$.

\section{Summary and Discussion}

In summary, by examining the quasiparticle excitation energy spectra and the gap equations, 
we have discussed the DCSB and the superconductivity 
in the generalized SNJL model.
We have found the finite-density effect in the critical coupling of the DCSB.
The effects of the bosonic part in the gap equation (18) in the SUSY BCS theory have been reviewed in detail,
while we have not found any SUSY effect 
in the thermodynamic character of the SUSY superconductivity.
These results have been understood by the energy spectra (10)-(13).
We have revealed that, the superconductivity shows the BCS character 
even if $\mu$ is close to $\Delta$.
There is no critical coupling also in the SUSY BCS case, 
while we have found the critical soft mass $\Delta^{sc(II)}_{cr}$:
SUSY protects not only the chiral invariance from DCSB but also
the dynamical breaking of gauge symmetry at finite density.
If the soft mass $\Delta$ is slightly larger than $\Delta^{sc(II)}_{cr}$, 
the SUSY effect becomes significant in the magnitude of the gap function $2|\phi_{S_{2}}|$.

Finally, we would like to make some comments on several issues and possible extensions of this work.
It is interesting for us to examine the collective modes and its excitations 
in both the DCSB and the superconductivity.
An interaction between bosons might alter the excitation energy spectra of the boson sector,
as discussed in the Bogoliubov theory of superfluidity~[22].

The existence of the upperbound of $\mu$ for the DCSB and superconductivity 
is a remarkable fact, when we consider phenomenological aspects or cosmological problems. 
Today, the superpartners of the known elementary particles are supposed
to exist in the TeV energy scale. From this point of view, any matters of massive Dirac particles
cannot take the chemical potential over the TeV region without destroying the superconductivity.
In this paper, we choosed the gauge freedom  broken by the superconductivity as $U(1)$ 
for the sake of simplicity. We regard our result is the starting point
to extend the theory to the more general $SU(N_{c})$ case.
From the phenomenologically possible energy scale of $\Delta$ ( $>$TeV ), 
our theory is relatively closer to the electroweak theory, 
especially the top condensation model~[8], or the technicolor theory~[7] than to $SU(3_{c})$ QCD.  
For example, the $SU(3_{c})$ gauge interaction itself is 
the origin of the attractive interaction for CSC. 
Therefore, if one uses an NJL-type model to describe CSC, 
the model parameters should be chosen from the consideration
of the QCD gauge interaction and hadron phenomenology. 
A coupling constant for an NJL-type model should take a numerical value of ${\cal O}({\rm GeV}^{-2})$, 
while a cutoff will become ${\cal O}({\rm GeV})$~[5,6].
At the order of the energy scale of the cutoff, there is no SUSY effect in CSC
in the sense of the context of this paper.

The applications of our method to an investigation of the SUSY superconductivity 
in (2+1)-dimensional case~[23] can easily be done.   
The gauge dynamics was completely neglected in this paper.
The gauged-NJL model was extensively studied in the context of the strong-coupling QED~[4].
Effects of gauge fields on the DCSB and superconductivity can be examined by our model 
at finite temperature and density. 
The extensions to several gauge groups of SUSY grand unified theories ( SGUTs ) are possible for our theory~[24].
Recently, the phase structures of the ordinary NJL model with an external electromagnetic field, 
both the cases in curved spacetime~[25] or at finite temperature and density~[26], are studied.
The DCSB of the $SU(N_{c})$ SNJL model at zero density was examined in several curved spacetime 
to describe the physics of the early universe. 
The extensions of our theory of this paper to these several external conditions are interesting~[27].

In this paper, we have given the possibility of the calculations of several thermodynamic quantities
( thermodynamic potential, entropy, heat capacity ) in a SUSY condensed matter.
It is one of the advantages of our method.
From our method, the Ginzburg-Landau theory for the DCSB and superconductivity can 
be constructed and applied to study vortex in the SUSY case.
Our theory provides a way to study the SUSY condensed matter physics.
The extension of our theory to the case of a vector-like $SU(N_{c})$-gauge model with $N_{f}$-flavor
will be published elsewhere~[28].

\begin{figure}

\caption{The excitation energy spectra of quasiparticles under the superconducting state with a finite chiral mass. 
We choose $\mu=2$, $|\phi_{S_{2}}|=0.02$, $\Delta=2$ and $|\phi_{S_{1}}|=1$.}

\caption{The critical coupling $(G_{1})_{cr}\Lambda^{2}$ of the DCSB, shown as a function of $\Delta/\Lambda$.}

\caption{The gap function $2|\phi_{S_{2}}|$ at $T=0$, shown as a function of $G_{2}$. 
We set $\mu=2$, $\Lambda=5$ and $|\phi_{S_{1}}|=1$.}

\caption{The gap function $2|\phi_{S_{2}}|$ at $T=0$, given as a function of $\Lambda$.
We set $\mu=2$, $G_{2}=0.5$ and $|\phi_{S_{1}}|=1$.}

\caption{The gap function $2|\phi_{S_{2}}|$ given as a function of $T$.
The model parameters are set as $\mu=2$, $\Lambda=5$, $G_{2}=0.5$ and $|\phi_{S_{1}}|=1$.
The curves, taken from the uppermost one, correspond to the values of
$\Delta=100$, $\Delta=10$, and $\Delta=2$ in the unit $|\phi_{S_{1}}|=1$.}

\end{figure}

\begin{table}
\caption{Relations of several orders.}
\begin{tabular}{cccc}
SUSY breaking mass & BEC & DCSB & CSC \\
\hline
$\Delta=\sqrt{\mu^{2}-|\phi_{1}|^{2}}$  & takes place & possible & no solution \\
$\Delta=\sqrt{\mu^{2}-4|\phi_{2}|^{2}}$ & takes place & no solution & possible \\
\end{tabular}
\end{table}

\end{document}

\title{Erratum: Dynamical Chiral Symmetry Breaking and Superconductivity 
in the Supersymmetric Nambu$-$Jona-Lasinio Model at finite Temperature and Density}
\author{Tadafumi Ohsaku}
\affiliation{Institute for Theoretical Physics, University of Cologne, Germany}

\date{\today}

%\maketitle

\newcommand{\bmx}{\mbox{\boldmath $x$}}
\newcommand{\bmy}{\mbox{\boldmath $y$}}
\newcommand{\bmk}{\mbox{\boldmath $k$}}
\newcommand{\bmp}{\mbox{\boldmath $p$}}
\newcommand{\bmq}{\mbox{\boldmath $q$}}
\newcommand{\bmP}{\mbox{\boldmath $P$}}  
\newcommand{\kfey}{\ooalign{\hfil/\hfil\crcr$k$}}
\newcommand{\pfey}{\ooalign{\hfil/\hfil\crcr$p$}}
\newcommand{\qfey}{\ooalign{\hfil/\hfil\crcr$q$}}
\newcommand{\Deltafey}{\ooalign{\hfil/\hfil\crcr$\Delta$}}
\newcommand{\nablafey}{\ooalign{\hfil/\hfil\crcr$\nabla$}}
\newcommand{\Dfey}{\ooalign{\hfil/\hfil\crcr$D$}}
\newcommand{\partfey}{\ooalign{\hfil/\hfil\crcr$\partial$}}
\def\sech{\mathop{\rm sech}\nolimits}

\begin{abstract}

Some equations of the paper, T. Ohsaku, {\it Dynamical Chiral Symmetry Breaking and Superconductivity 
in the Supersymmetric Nambu$-$Jona-Lasinio Model at finite Temperature and Density}, Phys. Lett. {\bf B634} (2006) 285-294, 
must be corrected. 
Fortunately, the numerical results of the original version of this paper are still valid
for our consideration on the supersymmetric BCS superconductivity.

\end{abstract}

\pacs{11.30.Pb, 11.30.Rd, 74.20.Fg}

\maketitle

First, several equations we must correct will be given bellow.
The matrix $\Omega_{F}$ of Eq. (7) in the original paper must be replaced by
\begin{eqnarray}
\Omega_{F} &\equiv& \left(
\begin{array}{cccc}
i\partfey+\gamma^{0}\mu-\frac{1}{2}\phi_{S_{1}}(1+i\gamma_{5})-\frac{1}{2}\phi^{\dagger}_{S_{1}}(1-i\gamma_{5}) & -\phi_{S_{3}}C(1+i\gamma_{5})-\phi^{\dagger}_{S_{2}}C(1-i\gamma_{5}) \\
-\phi_{S_{2}}C(1+i\gamma_{5})-\phi^{\dagger}_{S_{3}}C(1-i\gamma_{5}) & i\partfey^{T}-\gamma^{0T}\mu+\frac{1}{2}\phi_{S_{1}}(1+i\gamma_{5})+\frac{1}{2}\phi^{\dagger}_{S_{1}}(1-i\gamma_{5})
\end{array}
\right).   \nonumber
\end{eqnarray}
Here, the sign of the BCS-gap functions $\phi_{S_{2}}$ and $\phi_{S_{3}}$ of the block-off-diagonal part of 
$\Omega_{F}$ is different from that of the original version.
We derived the fermion quasiparticle excitation spectra and the fermionic contribution 
inside the self-consistent gap equations correctly in the original version,
while we must correct several formulae of the bosonic sector.
All of the places we mush correct are coming from this modification of signs of the gap function inside $\Omega_{F}$.

\vspace{3mm}

In order to take the symmetry of the BCS pairing gap in $\Omega_{F}$ as a Lorentz scalar,
we should choose now,
\begin{eqnarray}
\phi^{\dagger}_{S_{2}} = -\phi_{S_{3}}.  \nonumber
\end{eqnarray}
( We have chosen $\phi^{\dagger}_{S_{2}}=\phi_{S_{3}}$ in the original version. )
Under this choice, the bosonic quasiparticle energy spectra of Eqs. (10) and (11) of the original version will be replaced by 
\begin{eqnarray}
E^{B}_{+}(\bmp) &\equiv& \sqrt{ \bmp^{2}+|\phi_{S_{1}}|^{2}+4|\phi_{S_{2}}|^{2}+\Delta^{2}} - \mu,  \nonumber \\ 
E^{B}_{-}(\bmp) &\equiv& \sqrt{ \bmp^{2}+|\phi_{S_{1}}|^{2}+4|\phi_{S_{2}}|^{2}+\Delta^{2}} + \mu.  \nonumber 
\end{eqnarray}
Note that $E^{B}_{\pm}$ depend on $\bmp^{2}$ as monotonically incleasing functions.
Therefore, the gap equations Eqs. (17) and (18) of the original version are replaced as
\begin{eqnarray}
0 &=& \frac{\partial V_{eff}}{\partial |\phi_{S_{1}}| }  \nonumber \\
&=& \frac{2|\phi_{S_{1}}|}{G_{1}} - |\phi_{S_{1}}|\int^{\Lambda}\frac{d^{3}\bmp}{(2\pi)^{3}}\Bigl\{ \bigl( 1-\frac{\mu}{\sqrt{\bmp^{2}+|\phi_{S_{1}}|^{2} } }\bigr)\frac{1}{E^{F}_{+}}\tanh\frac{\beta}{2}E^{F}_{+}+\bigl( 1+\frac{\mu}{\sqrt{\bmp^{2}+|\phi_{S_{1}}|^{2} } }\bigr)\frac{1}{E^{F}_{-}}\tanh\frac{\beta}{2}E^{F}_{-} \Bigr\}  \nonumber \\
& & + |\phi_{S_{1}}|\int^{\Lambda}\frac{d^{3}\bmp}{(2\pi)^{3}}
\frac{1}{\sqrt{\bmp^{2}+|\phi_{S_{1}}|^{2}+4|\phi_{S_{2}}|^{2}+\Delta^{2}}}
\Bigl\{ \coth\frac{\beta}{2}E^{B}_{+} + \coth\frac{\beta}{2}E^{B}_{-}  \Bigr\},    \nonumber \\
0 &=& \frac{\partial V_{eff}}{\partial |\phi_{S_{2}}| }  \nonumber \\
&=& \frac{4|\phi_{S_{2}}|}{G_{2}} - 4|\phi_{S_{2}}|\int^{\Lambda}\frac{d^{3}\bmp}{(2\pi)^{3}}\Bigl\{ \frac{1}{E^{F}_{+}}\tanh\frac{\beta}{2}E^{F}_{+}+\frac{1}{E^{F}_{-}}\tanh\frac{\beta}{2}E^{F}_{-} \Bigr\}  \nonumber \\
& & + 4|\phi_{S_{2}}|\int^{\Lambda}\frac{d^{3}\bmp}{(2\pi)^{3}}
\frac{1}{\sqrt{\bmp^{2}+|\phi_{S_{1}}|^{2}+4|\phi_{S_{2}}|^{2}+\Delta^{2}}}
\Bigl\{ \coth\frac{\beta}{2}E^{B}_{+} + \coth\frac{\beta}{2}E^{B}_{-}  \Bigr\}, \nonumber 
\end{eqnarray}
respectively. Now, the charge density $\varrho$ given in Eq. (19) in the original paper becomes 
\begin{eqnarray}
\varrho &=& -\frac{\partial V_{eff}}{\partial \mu} \nonumber \\
&=& \int\frac{d^{3}\bmp}{(2\pi)^{3}} \Bigg( 
\frac{\mu-\sqrt{\bmp^{2}+|\phi_{S_{1}}|^{2}}}{E^{F}_{+}}\tanh\frac{\beta}{2}E^{F}_{+}+\frac{\mu+\sqrt{\bmp^{2}+|\phi_{S_{1}}|^{2}}}{E^{F}_{-}}\tanh\frac{\beta}{2}E^{F}_{-} +\coth\frac{\beta}{2}E^{B}_{+} - \coth\frac{\beta}{2}E^{B}_{-}   \Bigg). \nonumber   
\end{eqnarray}
The energy spectra $E^{B}_{\pm}$ at $\mu=0$ given in Eq. (21) of the original paper must be corrected into
\begin{eqnarray}
E^{B}_{\pm}(\bmp) &=& \sqrt{\bmp^{2}+|\phi_{S_{1}}|^{2}+4|\phi_{S_{2}}|+\Delta^{2}}.   \nonumber 
\end{eqnarray}
The solution of the equation $E^{B}_{+}(\bmp)=0$ given in Eq. (30) of the original paper will be replaced by
\begin{eqnarray}
|\bmp| &=& \sqrt{\mu^{2}-|\phi_{S_{1}}|^{2}-4|\phi_{S_{2}}|^{2}-\Delta^{2}}. \nonumber 
\end{eqnarray}

\vspace{3mm}

We give the condition on the Bose-Einstein condensation ( BEC ) of the scalar sector by the expression of $E^{B}_{+}$ given above.
For the positiveness of the Bose distribution function $1/(e^{\beta  E^{B}_{+}}-1)$,
\begin{eqnarray}
\sqrt{|\phi_{S_{1}}|^{2}+4|\phi_{S_{2}}|^{2}+\Delta^{2}} \ge \mu \ge 0  \nonumber 
\end{eqnarray}
must be hold.
The BEC takes place when $\mu=\sqrt{|\phi_{S_{1}}|^{2}+4|\phi_{S_{2}}|^{2}+\Delta^{2}}$ is satisfied.
This relation includes (27) and (29) of the original version.
Therefore, the existence of the upper bound of $\mu$ ( $\sim{\cal O}(\Delta)$ ) is true.
While, $\mu$ is the Fermi energy at the zero-temperature case, which we defined in the original version of this paper as follows:
\begin{eqnarray}
0 \le |\phi_{S_{1}}| \le \mu.   \nonumber 
\end{eqnarray}
Hence both of these inequality relations must be satisfied, as we revealed in the original version.

\vspace{3mm}

Finally, we wish to mention the numerical results given in the Sec. V of the original version.
( We do not have to correct the numerical result of Sec. IV in the original paper. ) 
The results are still valid. In Ref.~[1], the author discussed a similar system, 
a generalized Supersymmetric Nambu$-$Jona-Lasinio model with non-Abelian gauge/flavor degrees of freedom. 
The numerical results of Ref.~[1] ( solving the gap equations ) were performed under the following two cases:
(i) dynamical chiral symmetry breaking without color superconductivity,
(ii) color superconductivity with a vanishing chiral mass.
Hence, the numerical results of Ref.~[1] did not consider the situation where two orderings coexists
( i.e. the case $|\phi_{S_{1}}|\ne 0$, $|\phi_{S_{2}}|\ne 0$ ). 
In fact, errors inside the gap equations of the original version of this paper arises when we consider
the case of coexistence of two orderings.
Furthermore, the numerical results of the original version of this paper are qualitatively the same with that of Ref.~[1] .
The reason of the validity of the numerical results of the original version is that usually a contribution of the boson sector
in a gap function is small compared with that of the fermion sector.
( We confirm that numerical errors in Fig. 1 are ${\cal O}(10^{-3})$-${\cal O}(10^{-4})$, due to our choice $|\phi_{S_{2}}|/|\phi_{S_{1}}|=0.02$. 
In fact, all of branches $E^{B}_{\pm}$, $E^{F}_{\pm}$ have gaps, and several thermodynamic quantities are dominated by the contribution coming from the branch $E^{F}_{+}$ because the density of states of $E^{F}_{+}$ diverges at $|\bmk|=k_{F}$. ) 
Thus, numerical errors of the gap functions in the original paper are quite ( invisibly ) small.
Hence, we conclude that the BCS character of the supersymmetric superconductivity we obtained in the original version 
is also completely valid.